\documentclass[runningheads]{llncs}
\usepackage[T1]{fontenc}
\usepackage{graphicx,verbatim}
\usepackage{amsmath}
\usepackage{amsfonts}
\usepackage{booktabs}
\usepackage{array}
\usepackage{multirow}
\newcommand{\gap}[1]{\par\vskip #1\noindent}

\begin{document}
% \DeclareMathAccent{\vec}{\mathord}{letters}{"7E}
% %
\title{Search-MIND: Training-Free Multi-Modal Medical Image Registration}
\begin{comment}  %% Removed for anonymized MICCAI submission
\author{First Author\inst{1}\orcidID{0000-1111-2222-3333} \and
Second Author\inst{2,3}\orcidID{1111-2222-3333-4444} \and
Third Author\inst{3}\orcidID{2222--3333-4444-5555}}
%
\authorrunning{F. Author et al.}
\institute{Princeton University, Princeton NJ 08544, USA \and
Springer Heidelberg, Tiergartenstr. 17, 69121 Heidelberg, Germany
\email{lncs@springer.com}\\
\url{http://www.springer.com/gp/computer-science/lncs} \and
ABC Institute, Rupert-Karls-University Heidelberg, Heidelberg, Germany\\
\email{\{abc,lncs\}@uni-heidelberg.de}}

\end{comment}

% \author{******, ******,******, ****** and *****}  %% Added for anonymized MICCAI 2025 submission

\author{Boya Wang\inst{1}\thanks{Equal contribution}, Ruizhe Li\inst{1, 2, 3}\thanks{Equal contribution}, Chao Chen\inst{1, 2} and Xin Chen\inst{1}}  % Added for %anonymized MICCAI 2025 submission
\authorrunning{Boya Wang et al.}
\institute{
\inst{1}Intelligent Modelling \& Analysis Group (IMA),
\\\inst{2}Lab for Uncertainty in Data and Decision Making (LUCID),
\\School of Computer Science, University of Nottingham, UK
\\\inst{3} Nottingham Biomedical Research Centre (BRC), 
\\School of Medicine, University of Nottingham, UK}

\institute{
Intelligent Modelling \& Analysis Group (IMA), School of Computer Science, University of Nottingham, UK \and
Lab for Uncertainty in Data and Decision Making (LUCID), %School of Computer Science, University of Nottingham, UK %\and
Nottingham Biomedical Research Centre (BRC), School of Medicine, University of Nottingham, UK
}
% \institute{
% % 
% ****************
% }
\maketitle  % typeset the header of the contribution

% \author{Anonymized Authors}  %% Added for anonymized MICCAI submission
% \authorrunning{Anonymized Author et al.}
% \institute{Anonymized Affiliations \\
%     \email{email@anonymized.com}}
  
% \maketitle              % typeset the header of the contribution
%
\begin{abstract}
Multi-modal image registration plays a critical role in precision medicine but faces challenges from non-linear intensity relationships and local optima. While deep learning models enable rapid inference, they often suffer from generalization collapse on unseen modalities. To address this, we propose Search-MIND, a training-free, iterative optimization framework for instance-specific registration. Our pipeline utilizes a coarse-to-fine strategy: a hierarchical coarse alignment stage followed by deformable refinement. We introduce two novel loss functions: Variance-Weighted Mutual Information (VWMI), which prioritizes informative tissue regions to shield global alignment from background noise and uniform regions, and Search-MIND (S-MIND), which broadens the convergence basin of structural descriptors by considering larger local search range. Evaluations on CARE Liver 2025 and CHAOS Challenge datasets show that Search-MIND consistently outperforms classical baselines like ANTs and foundation model-based approaches like DINO-reg, offering superior stability across diverse modalities. 
\keywords{medical image registration \and training free  \and instance specific \and multi-modal \and weighted mutual information \and MIND descriptor}
% Authors must provide keywords and are not allowed to remove this Keyword section.

\end{abstract}
\section{Introduction}

Multi-modal image registration—the process of aligning images from disparate acquisition sources such as MRI, CT, and PET into a unified coordinate system—is a cornerstone of modern precision medicine~\cite{maintz1998survey,sotiras2013deformable}. In applications like image-guided radiotherapy and neurosurgery, this fusion is critical for mapping metabolic hotspots from PET or high-contrast soft tissue boundaries from MRI onto the geometrically accurate bone-referenced frames of CT~\cite{hill2001medical}. Beyond cross-modal fusion, the alignment of multi-sequence or multi-phase MRI (e.g., T1-weighted, T2-weighted, DWI, and contrast-enhanced phases) is essential for longitudinal disease monitoring and comprehensive tissue characterization~\cite{klein2009evaluation,modat2010fast}.

The fundamental challenge in these tasks lies in the non-linear intensity relationship between sensors; unlike mono-modal registration, where pixel intensities are highly correlated, multi-modal pairs often exhibit intensity inversions or entirely different physical representations of the same underlying anatomy~\cite{wells1996multi}. Consequently, successful registration requires moving beyond simple voxel-wise comparisons toward structural or information-theoretic descriptors that remain invariant to the specific imaging physics of each modality~\cite{heinrich2012mind}.

Current image registration paradigms are fundamentally divided into two categories: iterative optimization-based methods and learning-based frameworks. Iterative optimization treats registration as a latent variable estimation problem solved uniquely for each image pair, like ANTs (SyN)~\cite{avants2011reproducible} and Demons~\cite{wang2005validation}. While this approach provides high instance-specific precision without the need for prior training, it is inherently limited by three bottlenecks: (1) a high sensitivity to initialization and parameter setting, particularly under large rotations or scaling; (2) high computational cost when optimizing transformation parameters, particularly for deformable registration; and (3) the susceptibility of similarity metrics—such as the Modality Independent Neighbourhood Descriptor (MIND) —to converge toward local optima in the presence of noisy intensity distribution across different image modalities~\cite{heinrich2012mind}.

More recently, learning-based methods shift the computational burden to a pre-training phase. Task-specific models, such as VoxelMorph~\cite{balakrishnan2019voxelmorph}, enable rapid inference via learned displacement fields but often suffer from generalization collapse when applied to unseen patient populations or novel modalities. TransMorph~\cite{chen2022transmorph} introduces self-attention mechanisms to enhance global anatomical alignment; however, multimodal generalization remains limited by the training distribution. DiffusionReg~\cite{jiang2023se} models deformable registration as a diffusion-based generative process, iteratively refining deformation fields through denoising steps to capture complex, multimodal anatomical correspondences. More recently, foundation model-based approaches like DINO-reg~\cite{song2024dino} and SAMReg~\cite{huang2024samreg} have emerged, leveraging self-supervised Transformers like DINOv2~\cite{oquab2023dinov2} and SAM~\cite{kirillov2023segment} for feature representation. However, these generic features often lack the fine-grained structural invariance required to bridge the drastic appearance gap between CT and MRI, frequently failing to maintain the precise anatomical correspondence that classical optimization-based methods inherently preserve.

To address these challenges, we propose a training-free, iterative optimization framework for high-precision, instance-specific multi-modal image registration. By optimizing the transformation parameters directly at inference time, our approach avoids the generalization collapse typical of deep learning models when encountering novel modalities or unseen patient populations. The framework employs a two-stage pipeline: a hierarchical rigid-affine stage to resolve global rotation, translation, and scaling, followed by a deformable stage for non-rigid refinement. To mitigate sensitivity to initialization, we utilize a coarse-to-fine strategy that progressively narrows the search space toward global alignment. Central to this framework is the Search-MIND loss, a transformative metric that expands the original optimization search radius based on MIND descriptor~\cite{heinrich2012mind}, effectively overcoming the local optima that hinder conventional intensity-based methods. Our main contributions are three-fold:

\begin{enumerate}
  \item Universal Registration Paradigm: We present a domain-agnostic framework that enables immediate application to multi-patient and multi-modality data, bypassing the requirement for large-scale datasets, extensive pre-training and case-specific parameter tuning.
  \item Variance-Weighted Mutual Information (VWMI): We propose a differentiable, VWMI loss designed for coarse cross-modal alignment. By leveraging local intensity statistics, this metric adaptively prioritizes anatomically heterogeneous regions, effectively shielding the optimization process from the biasing effects of uniform background noise and uninformative padding.
  \item Search-MIND Loss (S-MIND): We propose a novel loss function that significantly broadens the convergence basin of the original MIND structural descriptors. This allows for stable and accurate alignment even in the presence of severe modality-specific artifacts and large anatomical deformations.
\end{enumerate}

\section{Method}

\subsection{Overview}
\begin{figure}[]
    \includegraphics[width=\textwidth]{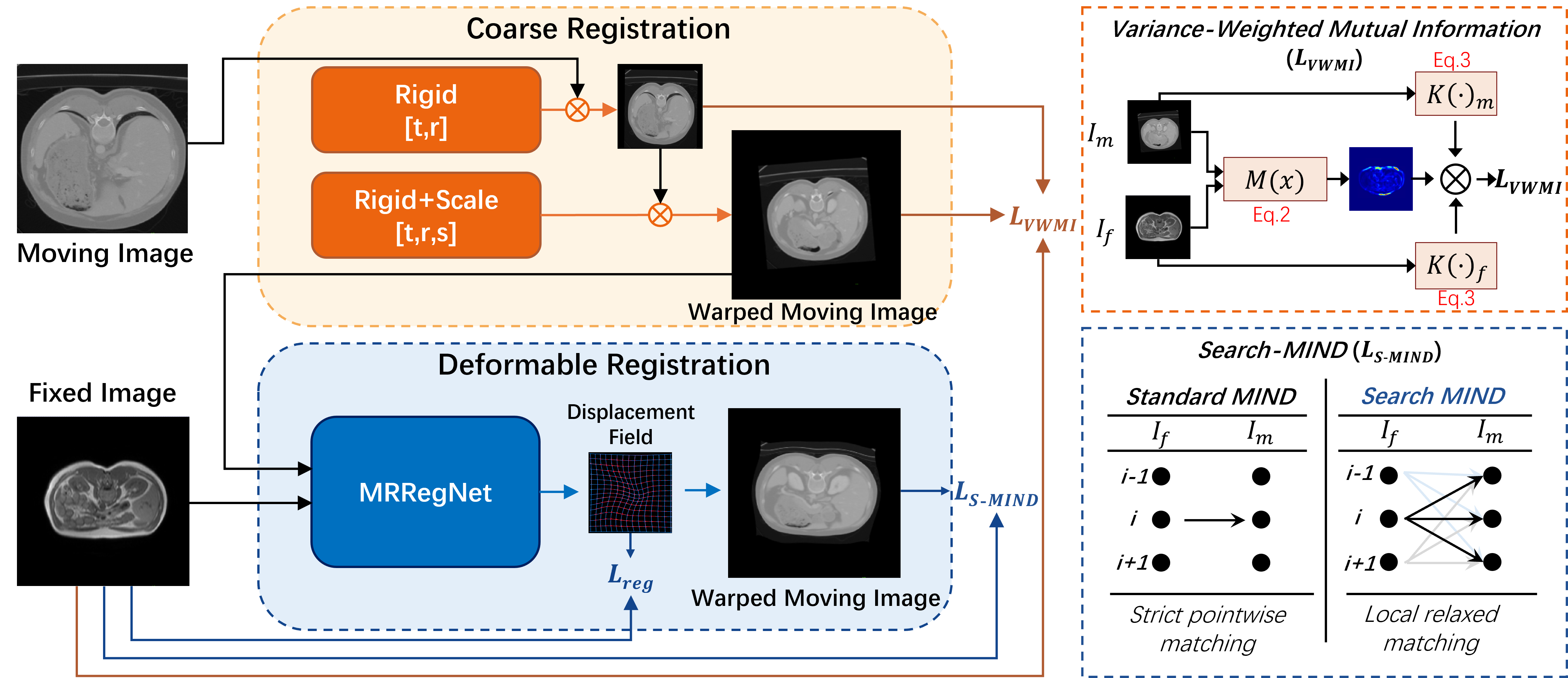}
    \caption{Overview of the proposed registration framework.} 
    \label{fig:framework}
\end{figure}

An overview of the proposed framework is shown in Fig.~\ref{fig:framework}. The pipeline follows a coarse-to-fine strategy, consisting of a two-stage coarse alignment followed by a deformable registration stage.

The coarse registration stage follows a classical optimization-based pipeline using the proposed VWMI loss function ($L_{\text{VWMI}}$) and optimized by gradient descent. The deformable registration stage adopts MRRegNet~\cite{li2024mrregnet} to predict multi-resolution residual deformation fields, optimized with the proposed S-MIND loss ($L_{\text{S-MIND}}$) and a diffusion regularization term $L_{\text{reg}}$.
\gap{-7mm}
\gap{-12mm}

\subsection{Data Preprocessing and Spatial Normalization}

To ensure anatomical consistency across diverse patient cohorts and imaging protocols, a rigorous preprocessing pipeline is implemented to achieve spatial normalization. First, all volumetric data are resampled to a standardized physical resolution of 1.0 mm $\times$ 1.0 mm $\times$ 2.5 mm, eliminating the spatial ambiguity inherent in varying sensor configurations. Subsequently, the volumes are cropped or padded to a fixed coordinate grid of 256 $\times$ 256 $\times$ 48, which centers the primary anatomical structures within a predictable spatial domain. This normalization provides a stable geometric initialization, allowing the subsequent optimization stages to focus on fine-grained structural alignment rather than compensating for drastic disparities in scale or orientation.

% \gap{-2mm}
\subsection{Coarse Registration via Variance-Weighted Mutual Information}

Unlike conventional deep learning paradigms, our approach eschews pre-trained weights in favor of instance-specific, inference-time optimization. To navigate the non-convex 9-degree-of-freedom (9-DOF) search space $\theta = [r, t, s]$ —comprising three rotations, three translations, and three scalings —we implement a simple optimization strategy guided by a newly proposed VWMI loss. 

To reduce computational overhead and broaden the initial convergence basin, both the moving and fixed volumes ($I_m$ and $I_f$) are downsampled by a factor of 2. We first optimize translation and rotation parameters to achieve global alignment. Once stable, scaling parameters are integrated into the optimization using the original image resolution.

% Traditional Mutual Information often suffers from bias caused by uniform background regions (e.g., air or padding) and imaging noise, which can dilute the statistical significance of anatomical structures~\cite{maes2003medical}. To address this, a differentiable adaptive masking mechanism based on local variance to prioritize in

Traditional Mutual Information is sensitive to uniform background and noise, which reduces anatomical discrimination~\cite{maes2003medical}. We therefore introduce a differentiable variance-based adaptive mask to emphasize informative tissue regions.

We define a spatial weight map $\mathcal{M}$ to prioritize high-contrast anatomical regions. Let $\Omega(\mathbf{x})$ denote a local neighborhood centered at voxel $\mathbf{x}$. The local variance $\sigma^2$ for image $I$ is defined as:

\begin{equation}
\sigma^2_I(\mathbf{x}) = \frac{1}{|\Omega|} \sum_{\mathbf{y} \in \Omega(\mathbf{x})} \left( I(\mathbf{y}) - \mu_I(\mathbf{x}) \right)^2
\end{equation}
where $\mu_I(\mathbf{x})$ is the mean intensity of the local region with dimension of $7\times7\times7$ in our setting. 

To emphasize regions where anatomical information exist in both the moving and fixed volumes, we define the joint weight map $\mathcal{M}$ as the normalized geometric mean of the two local standard deviations from $I_m$ and $I_f$ respectively:
\begin{equation}
\mathcal{M}(\mathbf{x}) = \frac{\sqrt{\sigma_{I_f}(\mathbf{x}) \cdot \sigma_{I_m}(\mathbf{x})}}{\max_{\mathbf{x}'} \sqrt{\left( \sigma_{I_f}(\mathbf{x}') \cdot \sigma_{I_m}(\mathbf{x}') \right)}}
\end{equation}

Using a Parzen window density estimator with a differentiable Gaussian function $K(\cdot)$, the weighted joint probability $P(i, j)$ for intensity bins $i$ and $j$ is formulated as:

\begin{equation}
P(i, j) = \frac{\sum_{\mathbf{x} \in \mathcal{D}} \mathcal{M}(\mathbf{x}) \cdot K(i - I_f(\mathbf{x})) \cdot K(j - I_m(\mathbf{x}))}{\sum_{\mathbf{x} \in \mathcal{D}} \mathcal{M}(\mathbf{x})}
\end{equation}
where $\mathcal{D}$ represents the image domain. The weighted marginal probabilities $P_f(i)$ and $P_m(j)$ are obtained through summation over the joint distribution:

\begin{equation}
P_f(i) = \sum_{j} P(i, j) \quad \text{and} \quad P_m(j) = \sum_{i} P(i, j)
\end{equation}

% \begin{equation}
% \mathcal{L}_{WMI}(I_f, I_m; \mathcal{M}) =1- \sum_{i \in \mathcal{L}_f} \sum_{j \in \mathcal{L}_m} P(i, j) \log \left( \frac{P(i, j)}{P_f(i)P_m(j)} + \epsilon \right)
% \end{equation}
The final VWMI loss is expressed as:
\begin{equation}
\mathcal{L}_{\text{VWMI}}(I_f, I_m; \mathcal{M}) =1- \sum_{i \in \mathcal{L}_f} \sum_{j \in \mathcal{L}_m} P(i, j) \log \left( \frac{P(i, j)+\epsilon}{P_f(i)P_m(j)+ \epsilon} \right)
\end{equation}
where $\mathcal{L}_f$ and $\mathcal{L}_m$ are the sets of discrete intensity bins (set to 32 bins in our experiment) for the fixed and moving volumes, respectively, and $\epsilon$ is a small constant ($10^{-7}$) for numerical stability.

\subsection{Deformable Registration via Search-MIND}

Conventional similarity measures for medical image registration, such as the MIND~\cite{heinrich2012mind}, rely on point-wise comparisons between corresponding locations. While effective for small deformations, such local formulations often suffer from limited capture range, especially in multi-modal scenarios (e.g., CT--MRI or T1--T2), where the optimization may be trapped in poor local minima.

To address this limitation, a search-based extension of MIND (S-MIND) is introduced to explicitly consider local displacements within a predefined window, thereby enlarging the effective receptive field of the similarity measure.

Let $I_f$ denote the fixed image and $I_m \circ \phi$ the warped moving image under deformation field $\phi$. Their corresponding MIND features are defined as
\begin{equation}
F_f = \mathrm{MIND}(I_f), \quad F_w = \mathrm{MIND}(I_m \circ \phi).
\end{equation}

For each voxel $x$, instead of directly comparing $F_f(x)$ and $F_w(x)$, a local search is performed over a discrete displacement set
\begin{equation}
\mathcal{S} = \{-r, \ldots, r\}.
\end{equation}

The feature distance under a shift $s\in S$ is defined as
\begin{equation}
d_s(x) = \left\| F_f(x) - F_w(x + s) \right\|.
\end{equation}

To obtain a differentiable formulation, a softmin operation is applied over the search window:
\begin{align}
\ell_s(x) &= -\frac{d_s(x)}{\tau} - \frac{s^2}{2\sigma^2},\\
% \end{equation}
% \begin{equation}
p(s \mid x) &= \frac{\exp(\ell_s(x))}{\sum_{s' \in \mathcal{S}} \exp(\ell_{s'}(x))},
\end{align}
where $\tau > 0$ controls the sharpness of the distribution and $\sigma$ controls the strength of the centre bias.

The centre bias term $-\frac{s^2}{2\sigma^2}$ is critical for regularizing the search space. By penalizing large displacements, it suppresses ambiguous matches from weakly discriminative structures and constrains the optimization toward locally consistent solutions. This mechanism significantly enhances stability and prevents global drift while still permitting larger shifts when strongly supported by the data. The final expected matching cost is defined as:
\begin{equation}
\tilde{d}(x) = \sum_{s \in \mathcal{S}} p(s \mid x)\, d_s(x).
\end{equation}

To reduce computational complexity, the search is performed independently along each spatial axis:
$\tilde{d}_x(x)$, $\tilde{d}_y(x)$, $\tilde{d}_z(x)$. The final search-based similarity loss is defined as
\begin{equation}
\mathcal{L}_{\text{S-MIND}} = \frac{1}{3} \left(
\mathbb{E}_x[\tilde{d}_x(x)] +
\mathbb{E}_x[\tilde{d}_y(x)] +
\mathbb{E}_x[\tilde{d}_z(x)]
\right).
\end{equation}

%The proposed formulation enlarges the capture range by allowing local displacement exploration within a predefined window. Compared to standard MIND similarity, it provides more informative gradients in early optimization stages, particularly for multi-modal registration. 

%In addition, the proposed formulation implicitly introduces directional guidance. Unlike conventional similarity measures that only compare co-located features, the search-based formulation evaluates multiple candidate displacements and assigns higher weights to better-matching directions through the softmin operation. As a result, the gradients are biased towards more plausible displacement directions, providing a form of direction-aware optimization without explicitly modeling vector fields.

By enabling local displacement exploration within a predefined window, the proposed formulation extends the capture range and provides more informative gradients during early optimization, especially in multi-modal contexts. Unlike conventional measures that compare only co-located features, this search-based approach implicitly introduces directional guidance by assigning higher weights to better-matching candidates via the softmin operation. This biases gradients toward plausible directions, enabling direction-aware optimization without explicit vector field modeling.

To ensure deformation smoothness, we apply diffusion regularization and employ the multi-resolution strategy of MRRegNet~\cite{li2024mrregnet}, which progressively refines the field through residual updates across scales. Let $\{\boldsymbol{\phi}_{\text{res}}^{(l)}\}_{l=1}^{L}$ denote the residual deformation fields from coarse to fine levels. The loss is defined as:
\begin{equation}
\mathcal{L}_{\text{reg}} = \frac{1}{L} \sum_{l=1}^{L} \left\| \nabla \boldsymbol{\phi}_{\text{res}}^{(l)} \right\|_2^2,
\end{equation}
where $\nabla \boldsymbol{\phi}_{res}^{(l)}$ denotes the smoothness term, approximating the spatial gradients of the deformation field using finite differences between neighboring voxels.

The final loss for deformable registration is defined as follows:
\begin{equation}
\mathcal{L}_{\text{deform}} = \mathcal{L}_{\text{S-MIND}} + \lambda \mathcal{L}_{\text{reg}},
\end{equation}
where $\lambda$ balances the alignment accuracy and deformation smoothness.

\section{Experiments and Results}

\subsection{Datasets}
We evaluated our method on two public datasets:

CARE Liver 2025 (Track 4)~\cite{liu2025merit}: This dataset contains multi-parametric MRI scans (T1, T2, DWI, and four Gd-EOB-DTPA-enhanced phases) from 610 patients. We performed intra-subject registration to align T1, T2, and DWI sequences to the GED4 reference. Evaluation was conducted on 148 image pairs for each modality, using the challenge-provided liver masks as ground truth.

CHAOS Challenge~\cite{kavur2021chaos}: This dataset comprises abdominal CT and MRI scans. We utilized 8 pairs for intra-subject registration and 56 pairs for inter-subject registration (CT to T1-MRI). The provided liver masks served as the basis for quantitative evaluation.

%The experiments were conducted on two public datasets: one is the \textbf{ MICCAI CARE Liver 2025 Track 4} dataset~\cite{liu2025merit} which contains 7 multi-parametric MRI scans, including T1-weighted (T1), T2-weighted (T2), diffusion-weighted imaging (DWI) and 4  Gd-EOB-DTPA-enhanced dynamic phases (GED1–GED4), from 610 patients diagnosed with liver fibrosis. Intra-subject image registration were performed to align T1, T2, DWI sequences to the GED4 reference image. In total, 148 pairs of images for each of the T1-GED4, T2-GED4 and DWI-GED4 were used for evaluation. Based on the annotations obtained in the CARE Liver 2025 Track 4 Challenge task, all sequences of the liver mask were available and were treated as ground truth for quantitative evaluation.

%The other dataset is from the \textbf{CHAOS Challenge}~\cite{kavur2021chaos} which consists of images of abdominal CT and MRI from different subjects. In total, 8 pairs of CT and T1-MRI were used to evaluate intra-subject image registration, and 56 pairs of CT and T1-MRI were used to evaluate inter-subject image registration. The liver masks provided by CHAOS were also used for quantitative evaluation.  

\subsection{Experimental Settings}

All methods were evaluated under a consistent protocol. We compared our approach against ANTs~\cite{avants2011reproducible} (Rigid and SyN variants) using default mutual information (MI) settings in a multi-resolution framework, and DINO-reg~\cite{song2024dino}, which was fine-tuned with a smoothness weight of 20, a $48 \times 48$ feature resolution, and a learning rate of 1 for optimal performance.

For the ablation study, we compared our proposed Ours (S-MIND) configured with $r=4, \tau=0.05, \sigma=2$, against Ours (MIND) using standard similarity. Both followed a sequential optimization: a coarse stage (Adam, LR: 0.01, 500 iterations) and a deformable stage (Adam, LR: $0.0001$, 200 iterations and $lambda=1$), both utilizing early stopping. Additionally, we evaluated rigid registration variants using MI (Ours-R-MI) and VWMI (Ours-R-VWMI).

To quantitatively evaluate registration performance, we employed three key metrics: the Dice Similarity Coefficient (DSC), calculated between the warped and target liver masks using the estimated transformation parameters; the percentage of folding voxels ($J \leq 0$); and the standard deviation of the log-Jacobian determinant ($\sigma(\log J)$) as a measure of deformation field smoothness. Statistical significance (Wilcoxon signed-rank test) is denoted by $^*$ ($p < 0.05$, deformable vs. S-MIND) and $^\dagger$ ($p < 0.05$, rigid vs. VWMI).

\gap{-3mm}
\subsection{Results}

\begin{table*}[!thb]
\centering
\caption{
Quantitative results on the CARE dataset.
Methods are grouped into pre-alignment (Initial and after coarse alignment), classical deformable baselines (ANTs-SyN, DINO-reg), and the proposed approaches.
}
\label{tab:care}
\setlength{\tabcolsep}{2.2pt}
\renewcommand{\arraystretch}{1.0}
\scriptsize
\resizebox{\textwidth}{!}{
\begin{tabular}{>{\centering\arraybackslash}m{2.3cm} ccc ccc ccc c}
\toprule
\multirow{2}{*}{Method}
& \multicolumn{3}{c}{T1} & \multicolumn{3}{c}{T2} & \multicolumn{3}{c}{DWI} & \multirow{2}{*}{T(s)$\downarrow$} \\
\cmidrule(lr){2-4}\cmidrule(lr){5-7}\cmidrule(lr){8-10}
& DSC$\uparrow$ & $J\!\leq\!0$$\downarrow$ & $\sigma(\log J)$$\downarrow$
& DSC$\uparrow$ & $J\!\leq\!0$$\downarrow$ & $\sigma(\log J)$$\downarrow$
& DSC$\uparrow$ & $J\!\leq\!0$$\downarrow$ & $\sigma(\log J)$$\downarrow$
 \\
\midrule
Initial
& 0.818 & --    & --
& 0.642 & --    & --
& 0.666 & --    & --
& -- \\
\midrule
ANTs-Rigid
& \textbf{0.879} & -- & --
& 0.703 & -- & --
& 0.705 & -- & --
& 10.0 \\
Ours (R-MI)
& 0.870 & -- & --
& 0.646$^\dagger$ & -- & --
& 0.703$^\dagger$ & -- & --
& 5.4 \\
Ours (R-VWMI)
& 0.871 & -- & --
& \textbf{0.710} & -- & --
& \textbf{0.725} & -- & --
& 4.8 \\
\midrule
ANTs-SyN
& 0.908* & 0.000 & 0.223
& 0.601* & 0.000 & 0.274
& 0.672* & 0.000 & 0.300
& 55.2 \\
DINO-reg
& 0.757*   & 0.951    & 0.126
& 0.685*   & 0.845    & 0.119
& 0.690    & 0.758   & 0.115
& 102.7 \\
\midrule
Ours (MIND)
& 0.918$^{*}$ & 0.045 & 0.153
& 0.759$^{*}$ & 0.270 & 0.363
& 0.776$^{*}$ & 0.263 & 0.450
& 44.6 \\
Ours (S-MIND)
& \textbf{0.920} & 0.064 & 0.186
& \textbf{0.761} & 0.129 & 0.354
& \textbf{0.790} & 0.260 & 0.483
& 53.5 \\
\bottomrule
\end{tabular}
}
% \end{table*}
\gap{5mm}
% \begin{table}[!thb]
\centering
\caption{
Quantitative results on the CHAOS dataset.
Methods are grouped into pre-alignment (Initial, and after coarse alignment), classical deformable baselines (ANTs-SyN, DINO-reg), and the proposed approaches.}
\label{tab:chaos}
\setlength{\tabcolsep}{2.2pt}
\renewcommand{\arraystretch}{1.0}
\scriptsize
% % \resizebox{\columnwidth}{!}{
\begin{tabular}{>{\centering\arraybackslash}m{2.3cm} ccc ccc c}
\toprule
\multirow{2}{*}{Method}
& \multicolumn{3}{c}{Same-Patient} & \multicolumn{3}{c}{Cross-Patient} & \multirow{2}{*}{T(s)$\downarrow$} \\
\cmidrule(lr){2-4}\cmidrule(lr){5-7}
& DSC$\uparrow$ & $J\!\leq\!0$$\downarrow$ & $\sigma(\log J)$$\downarrow$
& DSC$\uparrow$ & $J\!\leq\!0$$\downarrow$ & $\sigma(\log J)$$\downarrow$
\\
\toprule
Initial
& 0.457 & --    & --
& 0.441 & --    & --
& -- \\
\midrule
ANTs-Rigid
& 0.446 & -- & --
& 0.367$^\dagger$ & -- & --
& 12.9 \\
Ours (R-MI)
& \textbf{0.633} & -- & --
& \textbf{0.597} & -- & --
& 5.4 \\
Ours (R-VWMI)
& 0.606 & -- & --
& 0.564 & -- & --
& 5.0 \\
\midrule
ANTs-SyN
& 0.591 & 0.000 & 0.203
& 0.601* & 0.000 & 0.180
& 55.9 \\
DINO-reg
& 0.586*    & 5.87   & 0.190
& 0.363*    & 5.88   & 0.191
& 103.3 \\
\midrule
Ours (MIND)
& 0.683$^{*}$ & 0.413 & 0.497
& 0.633$^{*}$ & 1.045 & 0.577
& 44.8 \\
Ours (S-MIND)
& \textbf{0.718} & 0.373 & 0.521
& \textbf{0.656} & 0.985 & 0.565
& 52.6 \\
\bottomrule
\end{tabular}
\gap{-5mm}
\end{table*}

Tables~\ref{tab:care} and \ref{tab:chaos} report quantitative results on the CARE and CHAOS datasets, respectively. The proposed methods achieve consistently strong performance, with VWMI improving rigid alignment for CARE and S-MIND providing the best overall DSC for both dataset, while maintaining competitive deformation quality compared to ANTs-SyN and DINO-reg, as indicated by low folding ratios ($J \leq 0$) and stable $\sigma(\log J)$. DINO-reg shows higher folding (e.g., $J \leq 0 = 0.951$ on CARE T1), whereas ANTs-SyN produces nearly folding-free deformations due to its symmetric diffeomorphic formulation.

% On the CARE dataset, the performance gap is relatively small, as ANTs-SyN is a strong baseline for intra-modality MRI registration. In the rigid stage, both MI and VWMI perform competitively. VWMI achieves comparable results to MI and shows improvements on T2 and DWI (0.710 vs 0.646 and 0.725 vs 0.703), while no significant difference is observed on T1, likely because T1 and the target are from the same modality, where MI is already effective.

On CARE, performance differences are modest due to the strength of ANTs-SyN in intra-modality MRI registration. In the rigid stage, MI and VWMI perform comparably, with VWMI improving results on T2 and DWI, while showing no significant difference on T1, where modality consistency favors MI.

In cross-modality (CT--MRI) settings, the proposed S-MIND shows clearer advantages, achieving higher DSC than ANTs-SyN and DINO-reg (e.g., 0.718 vs 0.591/0.586 in same-patient, 0.656 vs 0.601/0.363 in cross-patient), as it relies on structure-aware representations rather than raw intensity similarity, which is less reliable across modalities. In the rigid stage, MI remains competitive due to its sensitivity to global intensity statistics and large-scale transformations, which are particularly important in CT--MRI alignment (0.633 vs 0.606). In contrast, VWMI introduces spatial weighting that emphasizes local structures, which can reduce sensitivity to global scale differences, leading to slightly lower DSC, however, the difference is not statistically significant. In the same-patient setting, the limited sample size (8 pairs) makes statistical significance difficult to establish despite consistent trends.

In terms of efficiency, the proposed methods are substantially faster than ANTs-SyN and DINO-reg (e.g., $\sim$50s vs 55--103s) while achieving comparable or better accuracy, demonstrating a favorable trade-off between performance and computational cost.

\gap{-1.5mm}
\section{Conclusion}
\gap{-1.5mm}
This paper presents Search-MIND, a training-free optimization framework that
addresses the persistent challenges of generalization collapse and local optima in multi-modal registration. By integrating Variance-Weighted Mutual Information for noise-resistant global alignment and the S-MIND loss to broaden the structural matching search radius, our method achieves high-precision, instance-specific results. Evaluations on the CARE Liver 2025 and CHAOS Challenge datasets confirm that Search-MIND outperforms both classical baselines and foundation model-based approaches. Ultimately, our domain-agnostic approach provides a robust, clinically viable solution for multi-modal fusion without requiring extensive pre-training or large-scale datasets. Future work may explore the generality and scalability of search-based similarity within the proposed framework, particularly with alternative multi-modal feature representations.

\bibliographystyle{splncs04}
\bibliography{reference}


% \bibitem{maintz1998survey}
% Maintz, J.B.A., Viergever, M.A.: A survey of medical image registration. Medical image analysis \textbf{2}(1), 1--36 (1998)

% \end{thebibliography}
\end{document}